\documentclass{aa}  
\usepackage{graphicx}
\usepackage{txfonts}
\usepackage{natbib}
\bibpunct{(}{)}{;}{a}{}{,}
\usepackage{epsfig}

\def\ltsima{$\; \buildrel < \over \sim \;$}
\def\lsim{\lower.5ex\hbox{\ltsima}}
\def\gtsima{$\; \buildrel > \over \sim \;$}
\def\gsim{\lower.5ex\hbox{\gtsima}}
\begin{document}
   \title{The optical afterglows and host galaxies of three short/hard gamma-ray bursts\thanks{The results reported in this paper are based on observations carried 
   out at ESO telescopes under programmes Id 076.A-0392, 078.D-0809 and 080.A-0825.}}


   \author{
P. D'Avanzo\inst{1} \and D. Malesani\inst{2} \and S. Covino\inst{1} \and S. Piranomonte\inst{3} \and
A. Grazian\inst{3} \and D. Fugazza\inst{1} \and R. Margutti\inst{1,4} \and V. D'Elia\inst{3} \and L. A. Antonelli\inst{3} \and 
S. Campana\inst{1} \and G. Chincarini\inst{1,4} \and M. Della Valle\inst{5,6,7} \and F. Fiore\inst{3} \and
P. Goldoni\inst{8,9} \and J. Mao\inst{1} \and R. Perna\inst{10} \and R. Salvaterra\inst{1} \and L. Stella\inst{3} \and 
G. Stratta\inst{11} \and G. Tagliaferri\inst{1}
          }

   \offprints{P. D'Avanzo, paolo.davanzo@brera.inaf.it}

   \institute{
INAF / Osservatorio Astronomico di Brera, via Emilio Bianchi 46, I-23807, Merate (LC), Italy. \and
Dark Cosmology Centre, Niels Bohr Institute, University of Copenhagen, Juliane Maries vej 30, DK-2100 K\o{}benhavn \O, Denmark. \and
INAF / Osservatorio Astronomico di Roma, via di Frascati 33, I-00040, Monteporzio Catone (Roma), Italy. \and
Universit\`a degli Studi di Milano-Bicocca, Dipartimento di Fisica, piazza delle Scienze 3, I-20126 Milano, Italy. \and
INAF / Osservatorio Astronomico di Capodimonte, salita Moiariello 16, I-80131 Napoli, Italy. \and
European Southern Observatory, Karl Schwarzschild Strasse 2, D-85748 Garching bei M\"unchen, M\"unchen, Germany. \and
International Center for Relativistic Astrophysics, piazza della Repubblica 10, I-65122, Pescara, Italy. \and
APC, Laboratoire Astroparticule et Cosmologie, 10 rue A. Domon et L. Duquet, 75205 Paris Cedex 13, France. \and
Service d'Astrophysique, DSM/IRFU/SAp, CEA-Saclay, F-91191 Gif-sur-Yvette, France. \and
JILA and Department of Astrophysical and Planetary Sciences, University of Colorado, 440 UCB, Boulder, CO, 80304, USA. \and
ASI Science Data Center, via Galileo Galilei, 00044, Frascati, Italy.
             }

   \date{Received; accepted}

 
  \abstract 
  {Our knowledge of short gamma-ray bursts (GRBs) has significatively improved
  in the \textit{Swift} era. Rapid multiband observations from the largest
  ground-based observatories led to the discovery of the optical afterglows
  and  host galaxies of these events. In spite of these advancements, the
  number of short GRBs with  secure detections in the optical is still fairly
  small. Short  GRBs are commonly thought to originate from the merging of
  double compact object  binaries but direct evidence for this scenario is
  still missing.}
  {Optical observations of short GRBs allow us to measure redshifts, firmly
  identify host galaxies, characterize their properties, and accurately
  localize GRBs within them. Multiwavelength observations of GRB afterglows
  provide useful information on the emission mechanisms at work. These are
  all key issues that allow one to discriminate among different models of these
  elusive events.}  
  {We carried out photometric observations of the short/hard GRB\,051227, 
  GRB\,061006, and GRB\,071227 with the ESO-VLT starting from several hours
  after the explosion down to the host galaxy level several days later. For
  GRB\,061006 and GRB\,071227 we also obtained spectroscopic observations of
  the host galaxy. We compared the results obtained from our optical 
  observations with the available X-ray data of these bursts.}  
  {For all the three above bursts, we discovered optical afterglows and firmly
  identified their host galaxies. About half a day after the burst, the optical
  afterglows of GRB\,051227 and GRB\,061006 present a decay significatly
  steeper than in the X-rays. In the case of GRB\,051227, the optical decay is
  so steep that it likely indicates different emission mechanisms in  the two
  wavelength ranges. The three hosts are blue star forming galaxies at
  moderate redshifts and with metallicities comparable to the Solar one. The
  projected offsets of the optical afterglows from their host galaxy center
  span a wide range, but all afterglows lie within the light of their hosts
  and present evidence for local absorption in their X-ray spectra. We
  discuss our findings in light of the current models of short GRB
  progenitors.}  
  {}
   

   \maketitle
%

\section{Introduction}

Gamma-ray bursts (GRBs) are rapid, intense flashes of gamma-ray radiation. 
Their cosmological nature was first suggested by their isotropic distribution 
in the sky \citep{Hakkila94}, and spectacularly confirmed by the detection of
long-wavelength  (X-rays to radio ranges) afterglows \citep{Costa97,
vanParadijs97, Frail97, Bremer98, Heng08}. The final  demonstration was
provided by the measurement of cosmological redshifts \citep{Metzger97}. 
Nowadays, these sources can be monitored up to weeks and months after the 
explosion. Two classes of GRBs are known to date, with a classification based
on their duration and spectral hardness \citep{Kouveliotou93}. Short GRBs (with
a duration $\la 2$ s) are on average harder than long ones (typically lasting
10--100 s). Although our knowledge of GRBs has significantly improved in recent
years, optical and near-infrared (NIR) studies of the afterglows and host
galaxies of short bursts are still limited to a small sample (e.g. 
\citealt{Fox05,Hjorth05,Berger05b,CastroTirado05,Bloom06,Covino06,Levan06};
see \citealt{Nakar07} for a review). While it has been firmly established that
long GRBs (or at least a significant fraction of them) originate in
core-collapse supernova (SN) explosions, the nature of short GRB progenitors is
still under debate. This is due, at least in part, to the fact that short GRBs
are detected less frequently (about one tenth of the GRBs detected by
\textit{Swift} belong to the short class; Berger 2007)  and have fainter
afterglows \citep{Kann08}.  Short GRBs with known distances have
isotropic-equivalent energies of  $\sim 10^{48} \mbox{--} 10^{52}$~erg, on 
average two orders of magnitude  lower than those of long GRBs. Moreover, the
association of several short GRBs with galaxies with low star formation
\citep{Gehrels05,Berger05b,Malesani07} and the tight upper limits on underlying
SNe \citep{Hjorth05b,Fox05,Covino06,Kann08} rule out models involving massive 
stellar progenitors. While the engine powering the burst may be similar for 
the short and long category, namely a black hole surrounded by an accretion 
torus, these observations strongly indicate different stellar progenitors  (for
recent reviews, see \citealt{Lee07,Nakar07}).

The merger of a double neutron star (DNS) or a black hole/neutron star (BH-NS)
binary system is  currently the leading model. In such systems, the delay
between binary formation and merging is driven by the gravitational wave
inspiral time (strongly dependent on the initial system separation). Some
systems are thus expected to drift away from the star-forming regions in which
they formed, before merging takes place. Simulations \citep{Belczynski02,
Belczynski06} show that a large fraction of the merging events should take
place outside, or in the outskirts, of galaxies. The density of the diffuse
medium in these regions is expected to be low and give rise to fainter
afterglows, setting in at later times than those of long GRBs (e.g.
\citealt{Vietri00,Panaitescu01}). A much faster evolutionary channel has been
proposed \citep{Belczynski01, Perna02, Belczynski06},  leading to merging
in only $\sim 10^6 \mbox{--} 10^7$~yr, when most systems  are still immersed in
their star forming regions.  The above scenarios are based on ``primordial''
binaries, i.e. systems that were born as binaries. Given the relatively small
delay between formation and merging ($< 1$~Gyr), the redshift distribution of
short GRBs that results from them should broadly follow that of the star
formation, especially at low redshift. Alternatively, a sizeable fraction of
DNSs may form dynamically by binary exchange interactions in globular clusters
during their core collapse \citep{Grindlay06}. The resulting time delay between
star-formation and merging would be dominated by the cluster core-collapse time
and thus be comparable to the Hubble time \citep{Hopman06}.  Both formation
channels (primordial and dynamical) may be  required to account for the
redshift distribution of short GRBs detected by \textit{Swift}
\citep{Salvaterra08, Guetta08}. Also, some short GRBs may originate from
different progenitors (such as hyperflares from soft gamma-ray repeaters;
\citealt{Hurley05, Tanvir05, Frederiks07, Mazets08}) and display different
afterglows, if any.

Key issues that could help in discriminating between the different theoretical
scenarios summarized above are reliable redshift determinations, the study of
the host galaxies properties, and accurate measurements of the spatial offsets
between afterglows and host galaxy centers, over a sufficiently large sample of
events. To date, an associated host galaxy candidate has been found for about half 
of the \textit{Swift} short GRBs. In particular, almost all well localized
short GRBs ($< 5\arcsec$ error radius) have a candidate host galaxy inside
their position error circle, but only for a dozen events with an observed
optical afterglow could a firm GRB-galaxy association be established. 
Among those bursts with an optical (sub-arcsec) localization, only GRB\,061201, 
GRB\,070809, and GRB\,080503 currently lack a secure host identification down to 
$R \sim 25\mbox{--}28$ \citep{Stratta07, Perley08a, Perley08b}.

In this paper we present the results of a detailed optical and NIR multiband
study of three short bursts,  GRB\,051227, GRB\,061006 and GRB\,071227. Our
discovery of their optical afterglows, together with deep optical  imaging at
late times, enabled us to firmly associate a host galaxy with each event. In
Sect. 3, 4 and 5 we present the analysis of our observations and of the 
available X-ray data for these bursts. In Sect. 6 we
discuss the properties of the afterglows and host galaxies of these GRBs  and
compare them with current models for the progenitors of short bursts. 
Throughout the paper we assume a flat Friedman-Robertson-Walker cosmology with
$H_0 = 71$ km s$^{-1}$ Mpc$^{-1}$, $\Omega_{\rm m} = 0.27$ and $\Omega_\Lambda
= 0.73$. All errors are at the $90\%$ confidence level, unless stated
otherwise.


\section{Observations and data analysis}

We observed the fields of GRB\,051227, GRB\,061006 and GRB\,071227 with the ESO
Very large Telescope (VLT), using the FORS1, FORS2 and ISAAC instruments. All
nights were clear, with seeing in the $0.6\arcsec\mbox{--}1.3\arcsec$ range.
For GRB\,051227 we observed the optical afterglow ($R$-band imaging) and the
host galaxy ($BVRIJ$-band imaging). For GRB\,061006 we obtained $I$-band images
of the optical afterglow, and photometric ($BVRIJK$) and spectroscopic
observations of the associated host galaxy. We carried out $R$-band imaging of
the afterglow of GRB\,071227 and optical spectroscopy of the associated host
galaxy. In order to look for the possible rising of a supernova associated
to GRB\,071227, we monitored its host galaxy in the $R$-band until about 40
days after the burst. A complete log of our observations, together with the
results of our analysis, is given in Tables \ref{tab_log1}, \ref{tab_log2}
and \ref{tab_log3}. Image reduction was carried out following the standard
procedures: subtraction of an averaged bias frame and division by a normalized
flat frame. The photometric calibration was achieved by observing  Landolt
standard fields (on different nights). Aperture and PSF-matched photometry were
performed using SExtractor \citep{Bertin96} and DAOPhot within
ESO-MIDAS\footnote{\texttt{http://www.eso.org/projects/esomidas/}},
respectively. Astrometric solutions were computed against the USNO-B1.0
catalogue\footnote{\texttt{http://www.nofs.navy.mil/data/fchpix/}}. In addition
to ordinary photometry, we also carried out image subtraction with the ISIS
package \citep{Alard98, Alard00}, a useful tool to find and pinpoint variable
sources in the field, even when blended with other objects. 

Our spectra of the host galaxy of GRB\,061006 and GRB\,071227 were acquired 
with the 300V grism (11~\AA{} FWHM), covering the 4000--9000~\AA{} wavelength
range. We used in all cases a $1\arcsec{}$ slit, resulting in an effective
resolution of $R = 440$. The extraction of the spectrum was performed with the
IRAF\footnote{\textit{IRAF} is distributed by the National Optical Astronomy
Observatories, which are operated by the Association of the Universities for
Research in Astronomy, Inc., under cooperative agreement with the National
Science Foundation.} software package.  Wavelength and flux calibration of the
spectra were achieved using a helium-argon lamp and observing
spectrophotometric stars. For both GRB\,061006 and GRB\,071227, we accounted
for slit losses by matching the flux-calibrated spectra to our photometry,
which was possible through a simple rescaling (by a factor of 1.4 and 2.6,
respectively). This shows that the derived spectral shape is robust.

The X-ray light curves and spectra of the three afterglows were obtained starting from
Level 1 products (event lists) processed by the Swift Data Center at
NASA/GSFC\footnote{\texttt{http://heasarc.gsfc.nasa.gov/docs/swift/sdc/}}. 
They were further processed following the standard procedure using the latest
HEAsoft release\footnote{\texttt{http://heasarc.gsfc.nasa.gov/docs/software/lheasoft/}} 
at the time of writing (2009 Jan.). The conversion from count rates to flux density
was computed assuming an absorbed power-law spectral model. We refer to the
upcoming paper by Margutti et al. (2009, in preparation) for details about the X-ray light curves
and spectra extraction.

\section{GRB\,051227}

GRB\,051227 was discovered by the \textit{Swift} coded mask Burst Alert
Telescope (BAT) on 2005 Dec. 27, at $t_0 = 18$:07:16 UT. The BAT light curve
shows a multi-peak structure, with a duration of $T_{90} = 8.0 \pm 0.2$ s and
a fluence of $(2.3 \pm 0.3) \times 10^{-7}$ erg cm$^{-2}$ in the 15--150 keV
band; the event was initially classified as a long burst \citep{Barbier05,
Hullinger05}. Further analysis of the \textit{Swift}/BAT data revealed that the
spectral lag between the 25--50 and 100--350 keV energy bands is negligible and
that the overall shape of the BAT light curve of this burst is similar to that
of the short/hard burst GRB\,050724, with an initial hard spike followed
several seconds later by a softer tail. These results indicate that GRB\,051227
very likely belongs to the short-duration class \citep{Barthelmy05a}. The
initial peak of GRB\,051227 was also detected by HETE--2. A joint analysis of
the HETE--2/FREGATE and \textit{Swift}/BAT data revealed that GRB\,051227 is a
hard burst \citep{Sakamoto05}. \textit{Swift}/XRT began to observe the GRB
field 93 s after the BAT trigger and found a bright fading uncatalogued X-ray
source. The X-ray light curve displayed a power-law like decay with small flares 
superposed \citep{Beardmore05}. 
Optical ground-based observations, carried out at the ESO-VLT,
revealed the presence of a variable source inside the XRT error box, identified
as the optical afterglow of GRB\,051227 \citep{Malesani05a, Malesani05b,
Malesani05c, DAvanzo05}. This was confirmed through Gemini observations
\citep{Soderberg05}. Further Gemini observations carried out on 2005 Dec 30
revealed that the afterglow had faded below the host galaxy level
\citep{Berger05a}.
The afterglow was also detected in the $U$ band at the 4$\sigma$ level by
\textit{Swift}/UVOT with a magnitude $19.8\pm 0.3$, but not in the other
passbands \citep{Roming05}. No radio counterpart was detected
\citep{Frail05}.

\begin{table*}
\caption{VLT observation log for GRB\,051227. Magnitudes are in the Vega system and are not corrected for
Galactic absorption. Errors and upper limits are given at $1\sigma$ and $3\sigma$ confidence level respectively.
}
\centering
\begin{tabular}{ccccccc} \hline
Mean time          &  Exposure time             & Time since GRB & Seeing    &  Instrument  &  Magnitude          & Filter \\
(UT)               &  (s)                       & (days)         & (\arcsec) &              &                     &        \\ \hline
2005 Dec  28.23379 &  $10 \times 240$           &  0.47874   	 & 1.3       & VLT/FORS1    & $24.55 \pm 0.11$    & $R$    \\
2005 Dec  29.21398 &  $11 \times 300$           &  1.45803     	 & 1.1       & VLT/FORS1    & $25.59 \pm 0.18$    & $R$    \\
2005 Dec  31.25894 &  $18 \times 360$           &  3.50299     	 & 0.9       & VLT/FORS1    & $25.49 \pm 0.09$    & $R$    \\
2006 Jan  02.26486 &  $20 \times 360$           &  5.50891       & 1.0       & VLT/FORS1    & $25.73 \pm 0.11$    & $V$    \\
2006 Jan  06.27768 &  $12 \times 300$           &  9.52173     	 & 1.0       & VLT/FORS1    & $26.06 \pm 0.15$    & $B$    \\
2006 Jan  06.29822 &  $6  \times 300$           &  9.54227     	 & 1.0       & VLT/FORS2    & $25.10 \pm 0.40$    & $I$    \\
2006 Mar  30.07032 &  $30 \times 4 \times 30$   & 92.31437       & 1.0       & VLT/ISAAC    & $> 23.1 $           & $J$    \\ \hline
\end{tabular}
\label{tab_log1}
\end{table*}


\subsection{Results}

   \begin{figure}
   \centering
   \includegraphics[angle=0,width=\columnwidth]{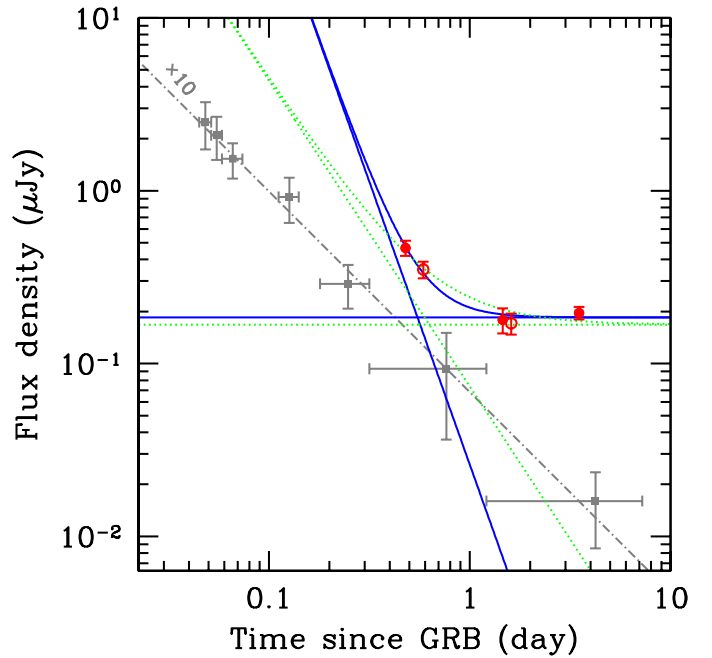}
      \caption{$R$-band light curve of the GRB\,051227 afterglow showing our VLT (filled circles) 
      and Gemini (empty circles; from \citealt{Bergeretal07a}) data. The solid and dotted lines show 
      the best power-law fit and the shallowest allowed decay (at 90\% confidence level), respectively. 
      The squares show the X-ray flux density at 1~keV (rescaled by a factor of 10), together with their best 
      power-law fit (dot-dashed line). See Sect. 3.1 for details.
	    }
       \label{fig:lc1}
   \end{figure}

\begin{figure*}
   \centering
   \includegraphics[width=\textwidth]{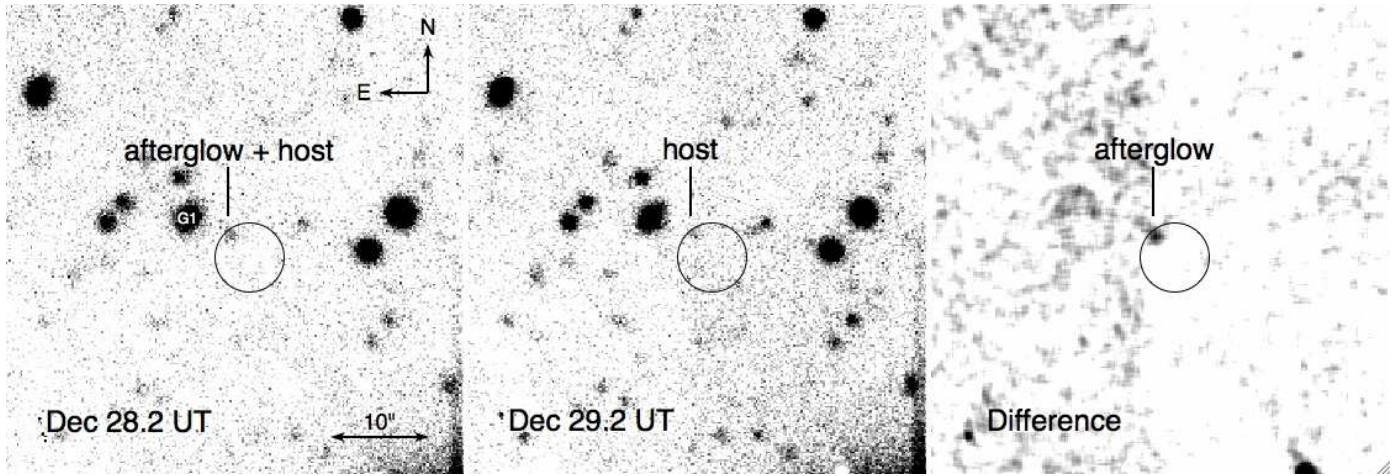}
      \caption{$R-$band images of the field of GRB\,051227 observed with the
      VLT about 0.5 (left) and 1.5 (middle) days after the burst. The fading
      of the afterglow is clearly visible in the subtraction image (right),
      which was smoothed for presentation purposes. The circle shows the X-ray
      position error \citep{Beardmore05}. The galaxy G1, identified by
      \citet{Foley05} and probably unrelated to the GRB, is marked in the left
      panel. See Sect. 3.1 for details.
	    }
       \label{fig:isis1}
   \end{figure*}

We started our observations of GRB\,051227 $\sim 0.5$ days after the burst,
soon after it was recognized as a short/hard burst (Hullinger et al. 2005;
Barthelmy et al. 2005a). In our first $R$-band image we found a point-like
object inside the XRT error circle (Malesani et al. 2005a). In order to look
for variability, we monitored the field at several epochs. Analysis of a second
$R$-band observation, carried out on 2005 Dec. 29, revealed that our candidate
had faded by $1.04 \pm 0.21$ mag with respect to our previous observation,
confirming the source as the optical afterglow of GRB\,051227 (Malesani et al.
2005b,c). Later $R$-band monitoring, carried out until Dec. 31, showed no
further variations of the brightness of the source, indicating that the host
galaxy contribution was dominant (see Table \ref{tab_log1} and Fig.
\ref{fig:lc1}). Given that we have only one detection of the afterglow in our
light curve, we can only place a lower limit onto the slope of the flux decay. 
Assuming a power-law decay $F(t) = F_0 + kt^{-\alpha}$, the decay index is
constrained to be $\alpha > 1.48$, and the host galaxy magnitude is $R =
25.52^{+0.18}_{-0.12}$. In order to better constrain the decay slope, we added
to our VLT data the Gemini-North observations reported by Berger et al.
(2007a), converting their magnitudes from the $r$ band by adopting the
conversion formulae recommended by the
SDSS\footnote{\texttt{http://www.sdss.org/dr6/algorithms/
sdssUBVRITransform.html\#Lupton2005}}. The combined best fit has $\alpha =
3.27^{+2.46}_{-1.50}$ and the host magnitude is $R = 25.55^{+0.16}_{-0.12}$
($\chi^2/\mathrm{d.o.f.} = 1.56/2$; Fig. \ref{fig:lc1}). 
An analysis of all the available \textit{Swift}/XRT data reveals an X$-$ray decay index significantly 
slower than the optical one around the same epoch. A power-law fit to the late points ($t > 2000$~s) 
of the 0.3--10 keV light curve provides  $\alpha_{\rm X} = 1.16^{+0.17}_{-0.25}$ 
($\chi^2/\mathrm{d.o.f.} = 1.16/5$; Fig.~\ref{fig:lc1}). An absorbed power-law fit to the X-ray 
spectrum gives a column density of $(2.3^{+1.1}_{-0.9}) \times 10^{21}$ cm$^{-2}$ 
($\chi^2/\mathrm{d.o.f.} = 18.92/22$), about a factor of 4 higher than the Galactic value 
in the direction of the burst.
Image subtraction performed with the ISIS package confirmed the decay of the afterglow 
between our first two epochs of VLT observations and the
flattening of the optical light curve (Fig. \ref{fig:isis1};
\citealt{DAvanzo05}). Moreover, perfoming astrometry on the residual image of
the subtraction process and on the images taken from  Dec. 29 onwards, we could
measure the accurate position of the afterglow inside the host galaxy. The
afterglow has coordinates (J2000) $\mbox{R.A.} = 08^{\rm h} 20^{\rm m}
58\fs10$, $\mbox{Decl.} = +31\degr 55\arcmin 31\farcs9$ ($0\farcs3$ error).
Its offset with respect to the galaxy center is $0\farcs2 \pm 0\farcs3$,
negligible within the errors, and consistent with the constraint set by
\citet{Bergeretal07a}. Using their limit (which is tighter than ours), and
remembering that the angular distance has a maximum at $z = 1.65$ for our
adopted cosmology, we infer a (projected) physical separation of less than 0.6~kpc
for any redshift. We note that the galaxy `G1' identified by \citet{Foley05} is
apparently not related to GRB\,051227.

We carried out further multiband photometry of the host galaxy in the period
2005 Dec. -- 2007 Mar., obtaining $B$-, $V$-, $R$- and $I$-band detections, and
a deep $J$-band upper limit. The results of our photometry are reported in
Table \ref{tab_log2}. The colors (corrected for Galactic extinction) are $B-V =
0.29 \pm 0.19$, $V-R = 0.16 \pm 0.23$ and $R-I = 0.41 \pm 0.45$. Due to the
faintness of the object, colors have large uncertainties, especially in the $I$
band, where detection is only marginal. Even with these large errors, the host
galaxy of GRB\,051227 appears very blue. 
Following \citet{Bergeretal07a}, we can use the apparent luminosity of the host 
galaxy as a rough indicator of its redshift. The host galaxy, with an 
extinction-corrected magnitude $R \sim 25.4$, is quite faint and this could 
suggest a moderately high redshift.
For comparison, we took a sample of galaxies with spectroscopic redshifts
from the Gemini Deep Deep Survey \citep{Abraham04}. We find that 90\% of the
systems in the magnitude range  $24 < R \la 25.5$ have redshift in the interval
$0.8 < z < 2.4$. The host galaxy  of GRB\,051227 is actually at the faint end
of the considered brightness range, hence it is probably at even higher
redshift. Conservatively, we infer $z \ga 0.8$ for this system. If real, the
UVOT detection of the afterglow in the $U$ 
filter\footnote{\texttt{http://www.swift.ac.uk/filters.shtml}} ($\lambda =
3500 \pm 500$~\AA) implies $z \la 2.8$, otherwise the observed  flux would be
suppressed by the intergalactic hydrogen. In addition we note that, within the
uncertainties of our photometry, the colors of the host galaxy are consistent
with those of an irregular galaxy at $z \sim 0.8$ \citep{Fukugita95}, in line
with our estimates.
At this redshift, the rest-frame $B$-band absolute  magnitude of the host would be 
$M_B = -17.2$ mag (or $L_B \sim 0.03L^*$, assuming the absolute $B$-band magnitude 
of field galaxies to be $M^*_B = -21$), and the GRB isotropic-equivalent energy would be
$E_{\rm \gamma,iso} = 3.9 \times 10^{50}$ erg (15--150 keV).

%

\section{GRB\,061006}

\textit{Swift} triggered GRB\,061006 on 2006 Oct 06 at $t_0 = 16$:45:50 UT. The
initial BAT light curve shows an impulse lasting about 5~s. The burst was
initially classified as long \citep{Schady06}. At the same time, RHESSI,
Konus/Wind and \textit{Suzaku} detected a short burst at the same coordinates,
with a duration $\leq 1$~s \citep{Hurley06}. A refined analysis of the BAT data
revealed that the burst began with an intense double spike at $t \sim t_0 -
22.5$~s followed by a softer persistent emission lasting at least until $t \sim
t_0 + 110$~s. The fluence in the 15--150 and 20--2000~keV bands was $(1.43 \pm
0.14) \times 10^{-6}$ and $3.6^{+0.3}_{-1.9} \times 10^{-6}$ erg~cm$^{-2}$,
respectively \citep{Krimm06,Golenetskii06}. \textit{Swift}/XRT began observing
143~s after the BAT trigger and found a faint fading X-ray source inside the
BAT error circle (Schady et al. 2006). The 0.3--10 keV light curve showed a 
broken power-law decay \citep{Troja06}.
No afterglow candidate was found with UVOT down to 18.5 mag (white light) in the prompt images at $t_0 +
140$~s (Schady et al. 2006). Optical ground-based observations, carried out at
the ESO-VLT 0.6 and 1.9 days after the burst, revealed the presence of a
variable source inside the XRT error box, identified as the optical afterglow
of GRB\,061006 \citep{Malesani06a, Malesani06b}.

\begin{table*}
\caption{VLT observation log for GRB\,061006. Magnitudes are in the Vega system and are not corrected for
Galactic absorption. Errors and upper limits are given at $1\sigma$ and $3\sigma$ confidence level respectively.}
\centering          
\begin{tabular}{ccccccc} \hline
Mean time         &  Exposure time             & Time since GRB & Seeing     &  Instrument  &  Magnitude          & Filter / grism \\
(UT)              &  (s)                       & (days)         & (\arcsec)  &              &                     &                \\ \hline
2006 Oct 07.32063 &  $10 \times 180$           &  0.62214  	& $0.8$      & VLT/FORS1    & $22.35 \pm 0.05$    & $I$ 	   \\
2006 Oct 08.30600 &  $10 \times 180$           &  1.91352  	& $0.8$      & VLT/FORS1    & $22.94 \pm 0.07$    & $I$ 	   \\
2006 Oct 09.19919 &  $7  \times 180$           &  2.50080  	& $0.9$      & VLT/FORS1    & $22.91 \pm 0.09$    & $I$ 	   \\
2006 Oct 08.77089 &  $180+300+165  $           &  2.07239  	& $0.9$      & VLT/FORS1    & $23.96 \pm 0.12$    & $R$ 	   \\
2007 Feb 17.65358 &  $40 \times 3 \times 20$   &  133.95509	& $1.2$      & VLT/ISAAC    & $> 21.2$            & $K$ 	   \\
2007 Feb 22.13000 &  $20 \times 180$           &  138.43153	& $0.6$      & VLT/FORS2    & $25.92 \pm 0.12$    & $B$ 	   \\
2007 Mar 10.07215 &  $47 \times 3 \times 20$   &  154.37368	& $0.8$      & VLT/ISAAC    & $22.00 \pm 0.20$    & $J$ 	   \\
2007 Mar 15.02346 &  $6  \times 180$           &  159.32499	& $0.9$      & VLT/FORS1    & $24.56 \pm 0.07$    & $V$ 	   \\
2007 Mar 15.03891 &  $3  \times 180$           &  159.34044	& $0.9$      & VLT/FORS1    & $23.05 \pm 0.12$    & $I$ 	   \\ \hline
2006 Oct 09.31165 &  $4 \times 2700$           &  2.613155      & $0.9$      & VLT/FORS1    & ---                 & 300V+GG375     \\ \hline
\end{tabular}
\label{tab_log2}
\end{table*}


\subsection{Results}

\begin{figure}
   \centering
   \includegraphics[width=\columnwidth]{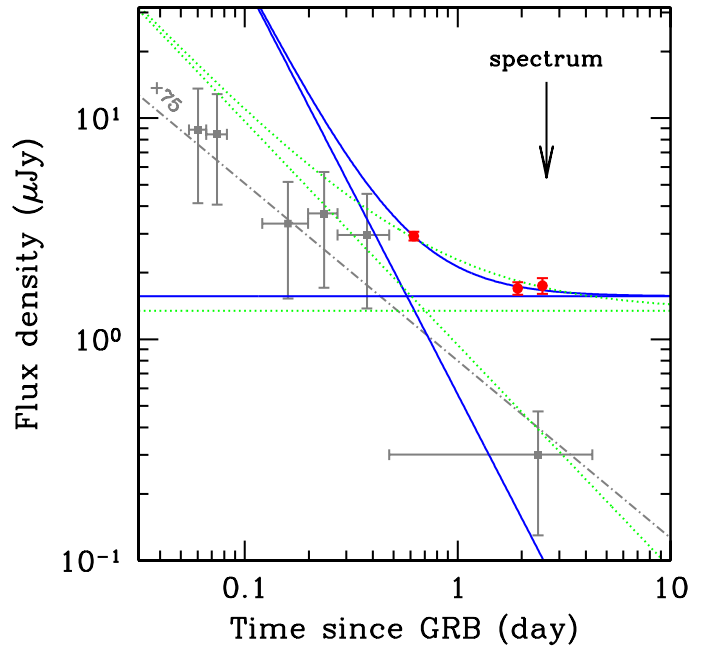}
      \caption{$I$-band light curve of the GRB\,061006 afterglow showing our VLT data 
      (filled circles). The solid and dotted lines show the best power-law fit 
      ($F(t) = F_0 + kt^{-\alpha}$) and the shallowest allowed decay, respectively. 
      The squares show the X-ray flux density at 1~keV (rescaled by a factor of 75), 
      together with their best power-law fit (dot-dashed line). The arrow indicates the 
      epoch of our spectrum. See Sect. 4.1 for details. 
      }
	\label{fig:lc2}
   \end{figure}

We started observing the field of GRB\,061006 about 0.6 days after the
burst, as soon as it became visible from the VLT. Given the rather high
Galactic reddening $E(B-V) = 0.32$ mag, we initially decided to adopt the
$I$ filter. In our first VLT image we found a point-like object inside the XRT
error circle (Troja et al. 2006b). To search for variability, we continued
monitoring the field at several epochs. Analysis of a second $I$-band
observation, carried out on 2006 Oct. 8 (i.e. 1.9 days after the burst),
revealed that the object inside the XRT error circle had faded by $0.59 \pm
0.09$ mag with respect to our previous observation, confirming that this source
was the optical afterglow of GRB\,061006 (Malesani et al. 2006a, 2006b). Later
$I$-band monitoring showed no further variations of the brightness of the
source, indicating that the afterglow had faded below the host galaxy level
(Table \ref{tab_log2} and Fig. \ref{fig:lc2}). Given that we have only one
detection  of the afterglow in our light curve, we can only set a lower limit
to the decay slope. Assuming a power-law like decay ($F(t) = F_0 +
kt^{-\alpha}$), the decay index is constrained to be $\alpha > 1.01$ (with a
best fit solution $\alpha = 1.86$; $\chi^2/\mathrm{d.o.f.} = 0.43/1$),
while the host galaxy magnitude is $I = 23.03^{+0.22}_{-0.15}$
(Fig.~\ref{fig:lc2}). Using the available X-rays data, we obtained the 0.3-10 keV 
light curve and spectrum of the afterglow. We performed a
power-law fit to the late points ($t > 300$~s) obtaining a  decay index of
$\alpha_{\rm X} = 0.80^{+0.13}_{-0.15}$ ($\chi^2/\mathrm{d.o.f.} = 1.74/7$). 
We note that the optical decay is steeper than the X-ray decay at the same epoch. 
The X-ray spectrum of the afterglow can be fitted with an absorbed power law with 
an absorbing column comparable to the (relatively large) Galactic value in the 
direction of the burst.

\begin{figure*}
   \centering
   \includegraphics[width=18.0cm]{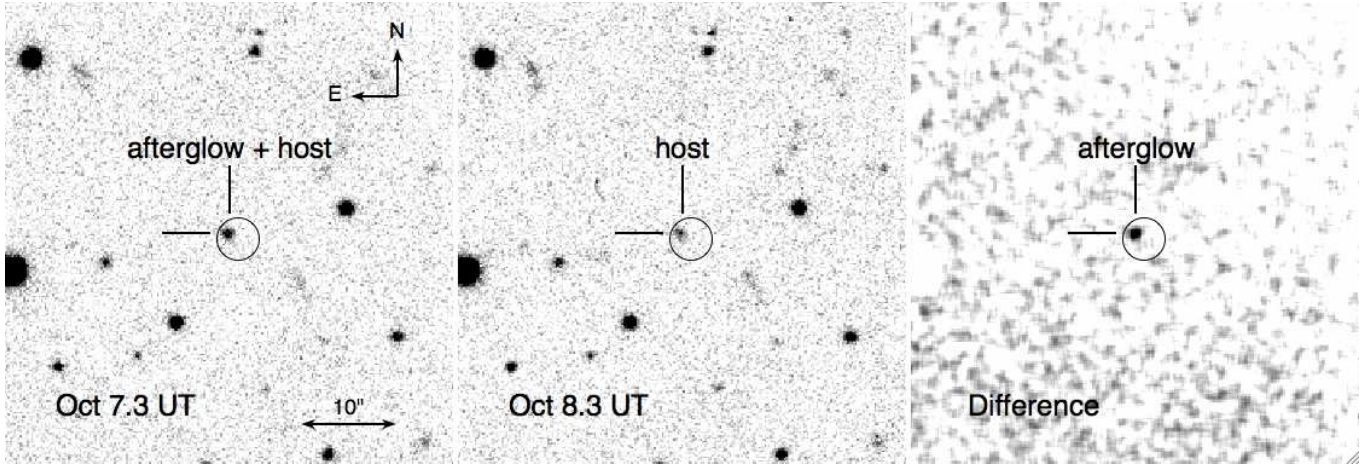}
      \caption{$I$-band images of the field of GRB\,061006 observed with the
      VLT about 0.6 (left) and 1.9 (middle) days after the burst. The fading
      of the afterglow is clearly visible in the subtraction image (right),
      which was smoothed for presentation purposes. The circle shows the X-ray
      position uncertainty \citep{Troja06b}. See Sect. 4.1 for details.
	    }
       \label{fig:isis2}
   \end{figure*}

Image subtraction performed with the ISIS package confirmed the decay of the
afterglow between the first two epochs (Fig. \ref{fig:isis2}) and the
flattening of the optical light curve. As in the case of GRB\,051227, we
measured the positions of the GRB\,061006 afterglow inside its host galaxy in
the residual image of the subtraction process. The optical afterglow
coordinates are (J2000): $\mbox{R.A.} = 07^{\rm h} 24^{\rm m} 07\fs60$,
$\mbox{Decl.} = -79\degr 11\arcmin 55\farcs1$ ($0\farcs2$ error). The
offset with respect to the host centre is $0\farcs3 \pm 0\farcs3$.

\begin{figure}
   \centering
   \includegraphics[angle=-90,width=8.0cm]{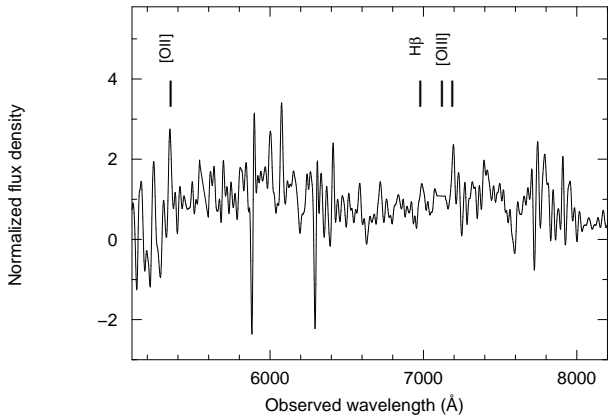}
      \caption{VLT/FORS1 spectrum of the host galaxy of GRB\,061006 at 
      $z~=~0.436~\pm~0.002$.
	     }
	\label{fig:spe}
   \end{figure}

On Oct. 9 we acquired a medium-resolution spectrum of the host galaxy. From the
detection of several emission lines ([O\,\textsc{ii}], H$\beta$, 
[O\,\textsc{iii}]), we derive a redshift $z = 0.436 \pm 0.002$ (Fig.
\ref{fig:spe}). This is consistent with the value reported by
\citet{Berger07,Berger08}. The resulting rest-frame $B$-band absolute magnitude
of the host is thus $M_B = -17.7$, or $L_B \sim 0.05L^*$  (assuming $M^*_B =
-21$). We secured further multiband photometry of the host galaxy in  the
period 2007 Feb. -- 2007 Mar., obtaining $B$, $V$, $R$, $I$ and $J$ detections,
together with a deep $K$-band upper limit. All the results of our photometry
are reported in Table~\ref{tab_log2}. The extinction-corrected colors are $B-V
= 1.04 \pm 0.14$, $V-R = 0.40 \pm 0.14$ and $R-I = 0.78 \pm 0.14$, which are
much bluer than those of elliptical galaxies and consistent with those of
spirals (Sbc or Scd type) at $z \sim 0.5$  (Fukugita et al. 1995). This is in
agreement with our spectroscopic redshift estimate. Considering the measured
angular offset, at $z = 0.436$ the afterglow is located within 3.5 kpc (in
projection) from the center of the host galaxy. At this redshift the
isotropic-equivalent energy of GRB\,061006 is $E_{\rm \gamma,iso} = 1.7 \times
10^{51}$ erg (20--2000 keV).

As indicated from several emission lines, star formation is still ongoing in
this galaxy. For the [O\,\textsc{ii}] emission line, we measure a luminosity of
$2.18 \times 10^{40}$ erg~s$^{-1}$ (corrected for slit losses). This converts
to an unobscured star formation rate (SFR) of $0.3~M_\odot$ yr$^{-1}$
\citep{Kennicutt98}, which corresponds to a specific SFR of about 
$6.1~M_\odot$ yr$^{-1} L_*^{-1}$. This is comparable to the typical  specific
SFRs measured in long GRB host galaxies \citep{Christensen04}.

The detection of oxygen and hydrogen emission lines in our spectrum allows us
to determine the metallicity of the host galaxy. This can be done by adopting
the  commonly used abundance diagnostic  $R_{23} =
(F_{[\mathrm{O}\,\mathsc{ii}] \lambda \,3727} +
F_{[\mathrm{O}\,\mathsc{iii}]\lambda \lambda \,4959,5007}) /
F_{\mathrm{H}\beta}$. However, as can be seen in Fig.~\ref{fig:spe}, the
H$\beta$ and [O\,\textsc{iii}]$\lambda$\,4959 emission lines have very low S/N
($\sim 2$), which makes the $R_{23}$ parameter of little practical use. To
overcome this problem, we used the empirical study performed by \citet{Nagao06}
on a large sample of galaxies, which shows that the ratio 
$F_{[\mathrm{O}\mathsc{iii}]\lambda \,5007} / F_{[\mathrm{O}\mathsc{ii}]\lambda
\,3727}$ can be used as a useful indicator of metallicity. For the host galaxy
of GRB\,061006 we obtain $F_{[\mathrm{O}\mathsc{iii}]\lambda \,5007} /
F_{[\mathrm{O}\mathsc{ii}]\lambda \,3727} = 0.52 \pm 0.23$  which corresponds,
according to the study of these authors, to an oxygen abundance range $8.2 \leq
12 + \log(\mathrm{O/H}) \leq 8.9$, i.e. a metallicity in the range
$(0.4\mbox{--}1.7) \times Z_\odot$ (using as solar abundance $12 +
\log(\mathrm{O/H}) = 8.69$  from \citealt{Allende01}). This is higher that what
commonly found for the host galaxies of long GRBs \citep{Savaglio08}.

%

\section{GRB\,071227}

\textit{Swift} triggered GRB\,071227 on 2007 Dec. 27 at $t_0 = 20$:13:47 UT. A
preliminary analysis of the BAT light curve revealed  the presence of at least
three pulses, with a total duration of about 5~s \citep{Sakamoto07}. From a
subsequent analysis of the BAT data in  the 15--350 keV range, the burst was
classified as short/hard, with a multipeaked light curve with $T_{90} = 1.8 \pm
0.4$~s and an extended emission up to $t = t_0 + 100$~s. The fluence in the
15--150 and 20--1300 keV bands was $(2.2 \pm 0.3) \times 10^{-7}$ and $(1.6 \pm
0.2) \times 10^{-6}$ erg~cm$^{-2}$, respectively \citep{Sato07,Golenetskii07}.
Konus/\textit{Wind} and \textit{Suzaku}/WAM detected the prompt emission of
GRB\,071227; the analysis of their data confirmed the short/hard nature of this
burst \citep{Golenetskii07, Onda08}. \textit{Swift}/XRT began observing the
GRB\,071227 field 80~s after the BAT trigger and found a fading, uncatalogued
X-ray source \citep{Sakamoto07}. The 0.3--10 keV light curve could be modelled
with a double broken power-law \citep{Beardmore07}.
\textit{Swift}/UVOT observations of the field of GRB\,071227, carried out
starting 86 s after the BAT trigger, revealed the presence of a source with a
white-filter magnitude of $21.7$ mag \citep{Sakamoto07,Cucchiara07}, which was
identified as an edge-on galaxy in ground-based observation with the Magellan
telescope \citep{Bergeretal07b}. Further observations with the ESO-VLT allowed
measurement of the galaxy redshift $z = 0.38$ and detection of a fading source at its
edge, identified as the optical afterglow of GRB\,071227
\citep{DAvanzo07,DAvanzo08}.

\begin{table*}
\caption{VLT observation log for GRB\,071227. Magnitudes are in the Vega system and are not corrected for
Galactic absorption. Magnitudes from Dec. 31.2 onwards are computed within an
annulus with a radius of about $1\farcs5$ centered on the afterglow position,
and are therefore not representative of the host galaxy. Errors are given at
a $1\sigma$ confidence level.}
\centering          
\begin{tabular}{ccccccc}
\hline
Mean time         &  Exposure time             & Time since GRB & Seeing     &  Instrument  &  Magnitude          & Filter / grism \\
(UT)              &  (s)                       & (days)         & (\arcsec)  &              &                     &                \\ \hline
\hline
2007 Dec 28.13346 &  $2 \times 120$            &   0.29055      & $0.7$      & VLT/FORS2    & $23.24 \pm 0.08$    & $R$          \\
2007 Dec 31.21576 &  $3 \times 180$            &   3.37133      & $0.7$      & VLT/FORS2    & $23.88 \pm 0.17$    & $R$          \\
2008 Jan 03.11081 &  $5 \times 180$            &   6.26790      & $0.8$      & VLT/FORS2    & $23.71 \pm 0.16$    & $R$          \\
2008 Jan 07.09081 &  $5 \times 180$            &  10.24790      & $0.8$      & VLT/FORS2    & $24.02 \pm 0.21$    & $R$          \\
2008 Jan 16.22578 &  $5 \times 180$            &  19.38287      & $0.7$      & VLT/FORS2    & $23.64 \pm 0.14$    & $R$          \\
2008 Jan 18.09430 &  $10 \times 180$           &  21.25139      & $0.8$      & VLT/FORS2    & $23.91 \pm 0.19$    & $R$          \\
2008 Jan 23.06066 &  $10 \times 180$           &  26.21775      & $0.9$      & VLT/FORS2    & $23.83 \pm 0.31$    & $R$          \\
2008 Feb 06.10412 &  $10 \times 180$           &  40.26121      & $1.1$      & VLT/FORS2    & $23.69 \pm 0.13$    & $R$          \\ \hline  
2007 Dec 28.21424 &  $4 \times 1800$           &   0.37133      & $0.7$      & VLT/FORS2    & ---                 & 300V+GG375   \\ \hline
\end{tabular}
\label{tab_log3}
\end{table*}

\subsection{Results}

   \begin{figure*}
   \centering
   \includegraphics[width=\textwidth]{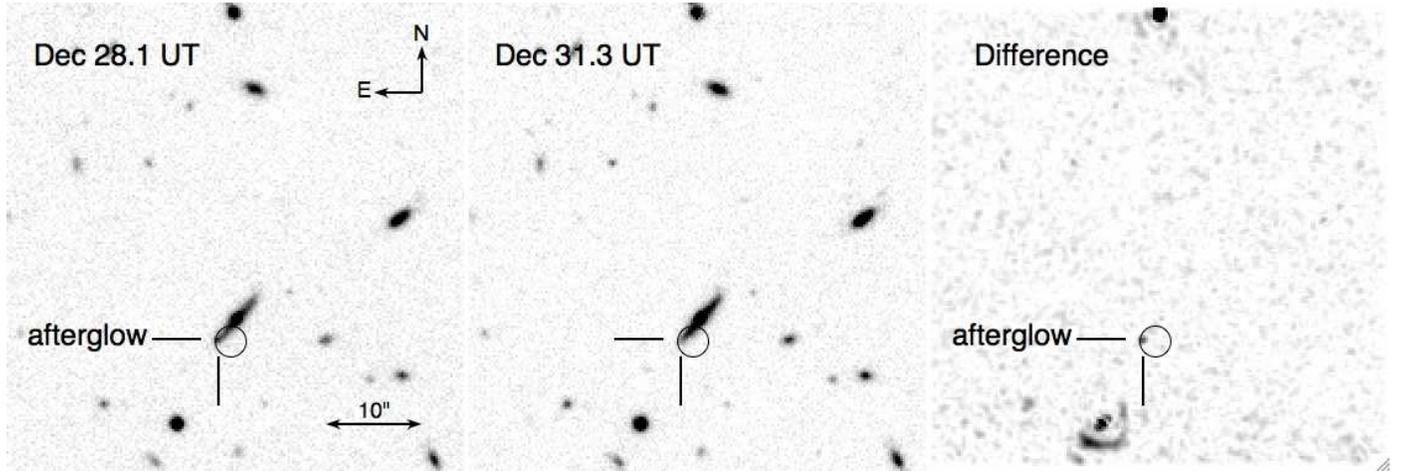}
      \caption{$R$-band images of the field of GRB\,071227 observed with the
      VLT about 0.3 (left) and 3.4 (middle) days after the burst.  The fading
      of the afterglow is clearly visible in the subtraction image (right). The
      circle shows the X-ray position error \citep{Beardmore07}. See Sect. 5.1
      for details.
	    }
       \label{fig:isis3}
   \end{figure*}

\begin{figure}
   \centering
   \includegraphics[angle=0,width=\columnwidth]{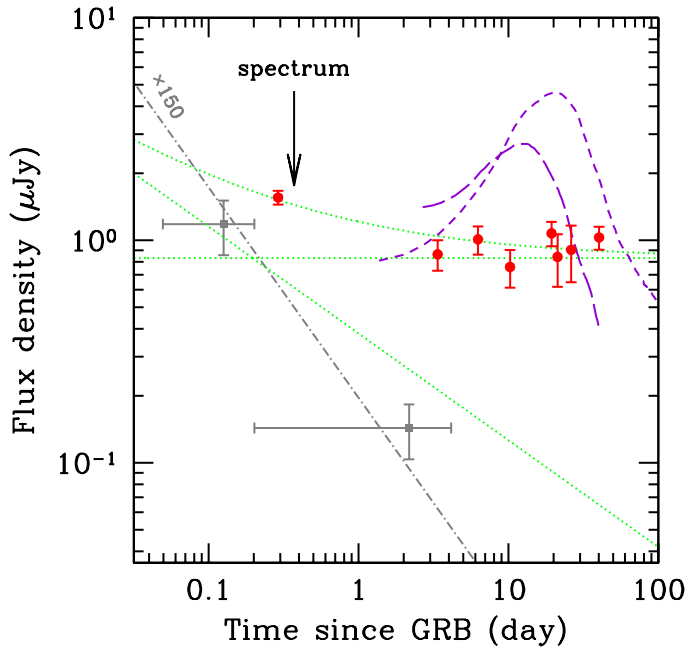}
      \caption{$R$-band light curve of the GRB\,071227 afterglow showing our VLT data 
      (filled circles). The dotted line shows the shallowest allowed decay. The squares show the X-ray flux density at 1~keV 
      (rescaled by a factor of 150), together with their best power-law fit (dot-dashed line). 
      The short- and long-dashed curves represent the light curves of SN\,1998bw and SN\,2006aj 
      \citep{Galama98,Pian06} reported at the redshift of GRB\,071227 ($z = 0.381$). The arrow 
      indicates the epoch of our spectrum. Note that the optical data are computed inside an 
      aperture of 1\farcs5 radius, and therefore they do not represent the host galaxy brightness. 
      See Sect. 5.1 for details.
      }
       \label{fig:lc3}
   \end{figure}

We observed the field of GRB\,071227 starting about 0.3 days after the burst.
Our first $R$-band image clearly showed the presence of a disk galaxy,
seen edge-on, overlapping with the XRT error circle, as already reported by
Cucchiara et al. (2007) and \citet{Bergeretal07b}. In addition, we also noted
the presence of a bright spot in the SE edge of the galaxy, consistent with the
XRT position. To check for variability, we took another image about 3 days
later. The result of image subtraction between the two epochs is shown in
Fig.~\ref{fig:isis3} and reveals that our candidate clearly faded
\citep{DAvanzo08}. The afterglow is superimposed on the plane of the host
galaxy and accurate photometry is thus challenging. However, the image
subtraction technique overcomes this problem. A ``reference frame'' (in this
case, our first-epoch image) is subtracted from all the  available images and
photometry is performed on the residual images. For any variable object in the
field the variation in flux, with respect  to the reference frame, is given as
output. Performing PSF photometry on the reference frame, it is possible to
calibrate in magnitude the flux  variations. The results of this procedure are
reported in Table~\ref{tab_log3}. Astrometry performed on the residual image of
the subtraction procedure gives the following coordinates (J2000): 
$\mbox{R.A.} = 03^{\rm h} 52^{\rm m} 31\fs23$, $\mbox{Decl.} = -55\degr
59\arcmin 03\farcs2$ ($0\farcs3$ error) for the optical afterglow, with a
projected offset of $2\farcs9 \pm 0\farcs4$  with respect to the host galaxy
center (located at, $\mbox{R.A.} = 03^{\rm h} 52^{\rm m} 31\fs02$,
$\mbox{Decl.} = -55\degr 59\arcmin 01\farcs0$). Further monitoring of the host
galaxy with the ESO-VLT showed no more variation in brightness up to about 40
days after the burst.  We measure a magnitude of $R \sim 20.6$ for the host
galaxy.
Assuming a power-law decay ($F(t) = F_0 + kt^{-\alpha}$), we can only
constrain the  decay index to be $\alpha > 0.48$. This is due to the fact that
we only have one datapoint during the afterglow phase. As expected for short bursts, 
in our light curve we find no evidence  for the presence of a SN simultaneous to the 
GRB, with a $3\sigma$ limit of $R = 24.9$ (Fig.~\ref{fig:lc3}).
Using the available X-ray data, we obtained the 0.3-10 keV 
light curve and spectrum of the afterglow. A double broken power-law fit gives a decay index of  
$\alpha_{\rm X} = 0.95^{+0.15}_{-0.16}$ ($\chi^2/\mathrm{d.o.f.} = 83.40/70$) around the 
epoch of our optical observations. The X-ray spectrum  of the afterglow can be fitted with 
an absorbed power law with an absorbing column of $(1.6^{+0.7}_{-0.6}) \times 10^{21}$ cm$^{-2}$, 
about 9 times higher than the Galactic value in that direction ($\chi^2/\mathrm{d.o.f.} = 41.28/48$).  

\begin{figure}
   \centering
   \includegraphics[angle=-90,width=8.0cm]{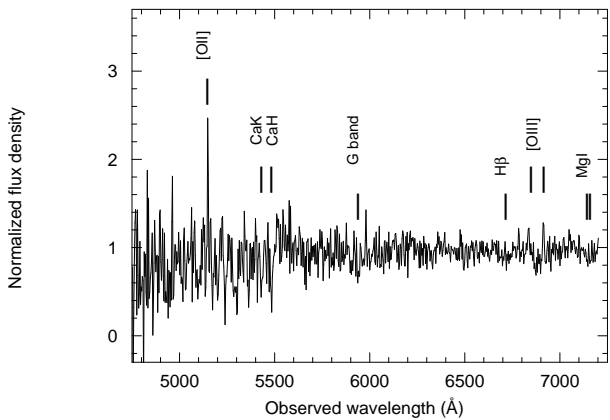}
      \caption{VLT/FORS2 spectrum of the host galaxy of GRB\,071227 at $z=0.381
      \pm 0.001$.
     }
	\label{fig:spe2}
   \end{figure}

On Dec. 27 we also carried out spectroscopic observations of the host galaxy of
GRB\,071227, with the medium-resolution grism 300V (Sect. 2). The slit was placed 
along the north-south direction and centered on the nucleus of the galaxy. 
From the detection of several emission ([O\,\textsc{ii}], [O\,\textsc{iii}] and weak H$\beta$) and
absorption lines we derived a redshift $z = 0.381 \pm 0.001$ (Fig.
\ref{fig:spe2}). This redshift was later confirmed by Berger et al. (2007). At
the measured redshift, the isotropic-equivalent energy of GRB\,071227 is
$E_{\rm \gamma,iso} = 5.8 \times 10^{50}$ erg (20--1300 keV),
a value typical of short GRBs. According to the offset  measured from our
astrometry, at this redshift the afterglow is located at $15.0 \pm 2.2$ kpc
from the center of the host galaxy. The host galaxy has a rest-frame $B$-band
absolute magnitude $M_B = -19.3$, i.e. $L_B \sim 0.2L^*$ (assuming $M^*_B =
-21$). The presence of emission lines in the optical spectrum of the host
galaxy of GRB\,071227 indicates that star formation is ongoing in this system.
We measure for the [O\,\textsc{ii}] emission line a luminosity of $3.9 \times
10^{40}$ erg s$^{-1}$ (corrected for slit losses). This translates into an 
unobscured SFR of 0.6 $M_{\odot}$ yr$^{-1}$ (Kennicutt 1998), which corresponds
to a specific  SFR of about 2.7 $M_{\odot}$ yr$^{-1} L_*^{-1}$. As we did
for the host of GRB\,061006 (Sect. 4.1), we can estimate the metallicity using
the $F_{[\mathrm{O}\mathsc{iii}]\lambda \,5007} /
F_{[\mathrm{O}\mathsc{ii}]\lambda \,3727}$ ratio as indicator \citep{Nagao06}. 
We obtain a value of $0.58 \pm 0.13$ which corresponds to an oxygen abundance
of $8.2 \leq 12 + \log(\mathrm{O/H}) \leq 8.8$, i.e. a metallicity in the range
$(0.4\mbox{--}1.0) \times Z_\odot$, using the solar abundance from
\citet{Allende01}. This estimate, similar to that obtained for the host galaxy
of GRB\,061006, is higher than those commonly found for the host galaxies of
long GRBs.

%

\section{Discussion}

The set of short GRBs with detected optical afterglow is still very small, and
the discovery of the optical counterparts for GRB\,051227, GRB\,061006 and GRB\,071227
provides new cases with secure host galaxy identification. Analysis of the host
properties provides substantial information on the progenitor systems.
Furthermore, the study of afterglow evolution can also be used to investigate
the emission processes of short GRBs.

\subsection{Afterglows}

Short GRB afterglow light curves have shown the same complexity observed in
long-duration ones, including flares, breaks and plateau phases (e.g.
\citealt{Soderberg06,deUgarte06,Roming06,Stratta07}). It is not always
straightforward to accomodate these features in the context of the standard
model, but multiwavelength observations are required to meaningfully test it. 
While X-ray light curves are available for many of the \textit{Swift} short
GRBs, the optical evolution is rarely well enough sampled, if at all. A
crucial issue is whether the emission in the two bands arises from the same
process.
The optical afterglows presented in this paper were too faint to
allow an extensive monitoring, hence we cannot say much about the temporal
evolution and the presence of breaks. However, both GRB\,051227 and GRB\,061006
present at $\sim 40\mbox{--}50$~ks a decay steeper in the optical than in the
X-rays. This is similar to what has been observed in some long GRBs and 
it cannot be straightforward explained within the standard external 
shock model, where this behaviour occurs only in the case of a wind-shaped
density profile. Such a medium is not expected around short GRB progenitors 
\citep{Nakar07, Lee07}.

The decay of the GRB\,051227 afterglow in the optical ($\alpha > 1.77$) is
actually particularly steep. Such indices are usually interpreted as evidence
of jetted emission. In this case, however, the X-ray light curve would be
expected to fall steeply as well, contrarily to what is observed. On the
other hand, very steep optical decays have been observed for the afterglows of
the short GRB\,050724 and GRB\,070707
\citep{Berger05b,Malesani07,Piranomonte08}. Also in these cases, the jet
interpretation has been rejected, and the light curve shape has been explained
assuming that long-lasting central engine activity was powering the afterglow
emission. In particular, for GRB\,050724 the optical emission was simultaneous
to a large X-ray flare \citep{Barthelmy05b,Campana06,Malesani07}, directly
supporting late-time activity from the engine. Only sparse X-ray data are
available for the INTEGRAL GRB\,070707 \citep{Piranomonte08,McGlynn08}, but
the observed optical decay index ($F(t) \propto (t-t_0)^{-\alpha}$) is too 
steep if computed assuming as the time origin the GRB trigger instant. The only
possible  explanation is long-term activity from the central engine, so that
the zero time  $t_0$ is moved to a later epoch, and the decay index is
shallower. We thus speculate that the rapid fading of GRB\,051227 (and possibly 
of GRB\,061006) could be due to an extra component visible in the
optical, but not in X-rays. However, as mentioned in Sect. 3.1, 4.1 and 5.1, none
of the GRBs presented in this work show flaring activity in their X-ray light
curves at the time of our optical observations.

\subsection{Host galaxies: star formation rates and metallicities}

The blue colors of the GRB\,051227 host suggest that active star formation is
ongoing in this system. In the cases of GRB\,061006 and GRB\,071227, the star
formation rate can be directly measured from the detection of [O\,\textsc{ii}]
in their optical spectra, at the level of about 6 and 3 $M_{\odot}$ yr$^{-1}
L_*^{-1}$, respectively. These values are comparable to the typical  SFR
observed in long GRB host galaxies (Christensen et al. 2004) and with those
reported by \citet{Berger08} for a sample of nine short GRBs with a secure host
galaxy association\footnote{With the significant exception of GRB\,050724 and
possibly GRB\,050509B, the only short GRBs firmly associated with early-type
galaxies and specific SFR lower than 0.01 and 0.03 $M_\odot$~yr$^{-1}$
${L_*}^{-1}$ \citep{Berger05b, Bloom06}.}. The presence of star forming
activity in many short GRB host galaxies does not exclude a core-collapse
origin for some of these events. With its relatively low redshift ($z = 0.381$)
and well resolved host galaxy (Fig.~\ref{fig:isis3}), GRB\,071227 is a perfect
testing ground for this hypothesis. To this end, we carried out an intensive
monitoring of the field of GRB\,071227 (Fig.~\ref{fig:lc3}) up to about 40 days
after the burst and we can exclude the presence of any associated supernova
down to $M_B > -15.1$. With such limit we can exclude the association of
GRB\,071227 with a bright type-Ib/c SN like those observed in long-duration
GRBs \citep{Galama98, Hjorth03, Stanek03, Malesani04, Pian06, WB06}. This agrees 
with other results from the literature \citep{Hjorth05, Covino06,
Kann08}. In a recent paper, \citet{Berger08} estimates metallicities for a
sample of five short GRB host galaxies, in the range $(0.6\mbox{--}1.6) \times
Z_\odot$. As reported in Sect. 4.1 and 5.1, we find similar values for the
metallicities of the host galaxies of GRB\,061006 and GRB\,071227. This shows
that short GRBs explode inside relatively evolved stellar populations, putting
an age constraint on the progenitor systems.

\subsection{Offsets}

   \begin{figure}
   \centering
   \includegraphics[angle=0,width=8.0cm]{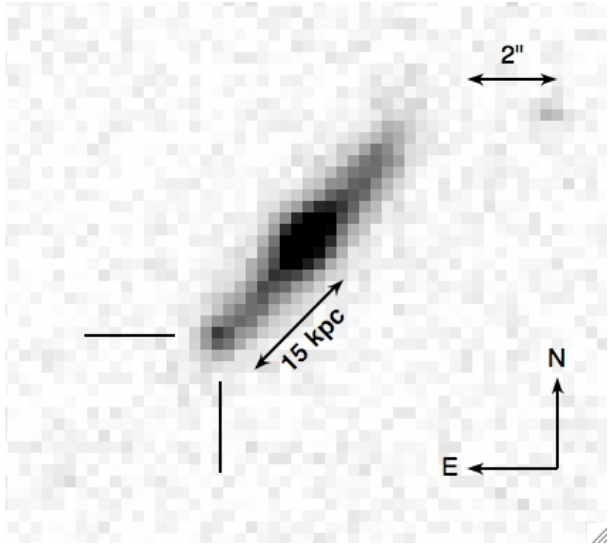}
      \caption{Closing on the optical afterglow of GRB\,071227 (marked by
      solid lines) and its host galaxy. The afterglow is located on top of the
      galaxy disk.
	    }
       \label{fig:GRB071227_detail}
   \end{figure}

As mentioned in Sect. 1, the study of the offset distribution of the afterglows
with respect to their host galaxies is a key issue in the study of short GRBs 
progenitors. In the context of double compact object progenitors, the offset
distribution contains information on the merging times and thus on the
evolutionary channels regulating binary systems evolution.

As reported in Sect. 5.1, the host galaxy of GRB\,071227 is a disk
galaxy,  seen edge-on and has a radius of about 15 kpc, like the Milky Way.
The GRB exploded on the galactic plane (Fig.~\ref{fig:GRB071227_detail}), where
probably most of the binary systems reside. Although we cannot exclude a chance
superposition, its localization does not argue for a binary system progenitor
born via binary exchange interactions in a globular cluster, given that most
globular clusters are located in the galactic halos. The progenitor of
GRB\,071227 might thus be an example of the merging of a ``primordial'' binary
system. The afterglow position of GRB\,071227 lies right in the middle of the
disc plane of its host galaxy, about 15 kpc in projection away from the bulge.
The large offset from the galaxy center could at first glance hint at a long
delay time between the formation of the binary system and its merging ($\sim
100$~Myr; \citealt{Belczynski02, Belczynski06}). In this case, the binary would
have  enough time to travel the required distance. However, the afterglow is
clearly lying on top of the galaxy light, hence any time delay can be
accomodated. Indeed, in case of an ejected system, it would be fortuitous that
the kick velocity vector was lying right in the galaxy plane.
The binary system that generated GRB\,071227 could indeed be a short-lived
one, with an evolutionary channel that foresees a merging in only $\sim 10^6
\mbox{--} 10^7$~yr \citep{Belczynski01, Perna02}, or it may have received a
negligible kick velocity. The system may thus still be embedded in its star
forming region. Concerning this last possibility, we note that the X-ray
spectrum of the afterglow of GRB\,071227 reveals significant metal absorption
(Sect. 5.1), indicating a dense medium.
Also, evidence for local absorption rules against a merging occurred in a
globular cluster, where the environment density is expected to be very low.
These considerations highlight once more the role that spectroscopy of 
short GRB afterglows plays in revealing the nature of the
progenitors of these events. 
Evidence for a dense medium is apparent also in the X-ray spectrum of the
afterglow of GRB\,051227 (Sect. 3.1). Even without an accurate redshift
determination, we can set a limit of 0.6~kpc on the projected offset between
the GRB explosion site and the host galaxy center (Sect. 3.1).  The host galaxy
of GRB\,061006 seems to be a star-forming galaxy too (Sect. 4.1), but less
luminous, and probably smaller, than the host galaxy of GRB\,071227. In our
VLT images we cannot spatially resolve the galaxy but, thanks to the image
subtraction technique, we are able to constrain the afterglow offset with
respect to its center, which is smaller than 3.5 kpc (in projection). No excess
column density has been found in the X-ray spectrum (Sect. 4.1), but the
limits are not very constraining due to the large foreground component ($1.3
\times 10^{21}$~cm$^{-2}$).
The small offsets and the presence of gas absorption point towards a 
``primordial'' binary progenitor also for GRB\,061006 and GRB\,051227. In these
cases too, an optical spectrum taken during the early stages of the afterglow
could have given useful hints. Featureless spectra have recently been reported
for the short bursts GRB\,061201 \citep{Stratta07} and GRB\,070707
\citep{Piranomonte08}, but with a low signal-to-noise.

Finally, we note that the three short GRBs studied in this paper show an
extended-duration soft emission component in their gamma-ray prompt light
curves \citep{Barthelmy05a, Krimm06, Sato07} and, as said above, all of them
are located within the light of their host galaxy. Recently, \citet{Troja08}
suggested that GRBs showing an extended-duration soft emission component preferentially 
have small offsets. This could be a consequence of a common physical
progenitor (BH-NS binary), expected to merge on short timescales, which would
lead to very little distance travelled before producing the GRB. Independently of 
its interpretation, such a conclusion has to be considered with some caution, given
that not all the short GRBs have a secure host galaxy association (and,
consequently, some of the offsets reported in the literature could be
misleading; \citealt{Piranomonte08}). 
Alternatively, as suggested by \citet{Bernardini07}, the presence of a
temporally long-lasting soft tail observed in short GRBs could be related to
the density of the circumburst medium.

\section{Conclusions}

We have presented the results of a multiband optical and NIR observational
campaign on the three short/hard GRBs GRB\,051227, GRB\,061006 and GRB\,071227.
We discovered their optical afterglows and identified their host galaxies. 

The optical light curves of GRB\,051227 and GRB\,061006 show evidence for a
steep decay between $\sim 0.5$ and $\sim 1.5$ days after the GRB, at variance
with the shallow decay of their X-ray afterglows measured at similar epochs.
These findings are difficult to explain based on the standard external
shock model and, particularly for the case of GRB\,051227, this behaviour
can be interpreted as due to  different mechanisms producing the optical and
X-ray afterglows.

With our dataset we were able to characterize the host galaxies of these
short GRBs. These galaxies are of late type, lie at relatively low redshifts
and present a moderate level of star formation activity. The absence of
brightening in the late optical light curve of GRB\,071227 rules against 
an association with a bright core collapse SN, though the limit $M_B > -15$ on
the peak absolute magnitude cannot exclude the presence of a core-collapse
explosion characterized by low energy and very little content of $^{56}$Ni in
the ejecta \citep{DellaValle06, Tominaga07}. The offsets between the
afterglows and the centers of their host galaxies may suggest that these short
GRBs originated from the merging of double compact objects system likely of
``primordial'' origin. In particular, the location of GRB\,071227 on the plane
of a disk galaxy argues against ejection. The detection of an absorbing
column in the X-ray spectrum may indeed indicate that the binary was not far
from its star forming region. Looking at the near future, monitoring of the
early stages of the afterglows of short GRBs, together with prompt optical
spectroscopy, will likely shed light  on many open issues on the origin of
these events.

%

\begin{acknowledgements}
We thank the referee for his/her useful comments and suggestions. 
PDA acknowledge the  Italian Space Agency for financial support through the
project ASI I/R/023/05. DM acknowledges the Instrument Center for Danish
Astrophysics for financial support. The Dark Cosmology Centre is funded by the
Danish National Research Foundation. We acknowledge the invaluable help from
the ESO staff at Paranal in carrying out our target-of-opportunity observations. 
This work made use of data supplied by the Swift Data Center at NASA/GSFC.
\end{acknowledgements}

\bibliographystyle{aa}
\bibliography{references}

\end{document}